\documentclass[12pt]{article}

\usepackage{amssymb,latexsym,amsfonts,fullpage,amsmath}

\input{Qcircuit}

\newcommand{\sI}{\mathrm{I}}
\newcommand{\sS}{\mathrm{S}}
\newcommand{\sW}{\mathrm{W}}
\newcommand{\sR}{\mathrm{R}}
\newcommand{\sA}{\mathrm{A}}

\newcommand{\sG}{\mathrm{G}}

\newcommand{\mI}{\mathbb{I}}
\newcommand{\mS}{\mathbb{S}}
\newcommand{\mW}{\mathbb{W}}
\newcommand{\mR}{\mathbb{R}}
\newcommand{\mA}{\mathbb{A}}

\newcommand{\mG}{\mathbb{G}}

\newcommand{\Z}{{\mathbb Z}}
\newcommand{\R}{{\mathbb R}}
\newcommand{\N}{{\mathbb N}}
\newcommand{\C}{{\mathbb C}}
\newcommand{\cH}{{\cal H}}

\newcommand{\cL}{{\cal L}}

\newcommand{\cP}{{\cal P}}
\newcommand{\cV}{{\cal V}}
\newcommand{\cS}{{\cal S}}

\newcommand{\<}{\langle}
\renewcommand{\>}{\rangle}
\newtheorem{Definition}{Definition}
\newtheorem{Lemma}{Lemma}
\newtheorem{Theorem}{Theorem}

\newcommand{\onemat}{{\bf 1}}

\newcommand{\lra}{\rightarrow}

\newcommand{\band}[2]{
\begin{array}{|cc|}
\hline
#1 & #2 \\
\hline
\end{array}
}

\newcommand{\doubleband}[4]{
\begin{array}{|cc|}
\hline
#1 & #2 \\
#3 & #4 \\
\hline
\end{array}
}

\newcommand{\W}{\mathrm{W}}
\newcommand{\G}{\mathrm{G}}
\newcommand{\F}{\mathrm{F}}

\newcommand{\bG}{\mathbb{G}}
\newcommand{\bF}{\mathbb{F}}

\title{A single-shot measurement of the energy of product states in a translation invariant spin chain can replace any quantum computation}

\author{Dominik Janzing\thanks{School of Electrical Engineering and Computer Science, University of Central Florida, Orlando, FL 32816, USA. Email: \texttt{janzing@ira.uka.de}}\quad 
Pawel Wocjan\thanks{School of Electrical Engineering and Computer Science, University of Central Florida, Orlando, FL 32816, USA, Email: \texttt{wocjan@cs.ucf.edu}}\quad
Shengyu Zhang\thanks{Institute for Quantum Information, California Institute of Technology, Pasadena, CA 91125, USA. Email: \texttt{szhang@cs.caltech.edu}}}

\begin{document}
\date{December 14, 2007}
\maketitle

\abstract{In measurement-based quantum computation, quantum algorithms are implemented via {\it sequences} of measurements.  We describe a translationally invariant finite-range interaction on a one-dimensional qudit chain and prove that a {\it single-shot} measurement of the {\it energy} of an appropriate computational basis state with respect to this Hamiltonian provides the output of any quantum circuit. The required measurement accuracy scales inverse polynomially with the size of the simulated quantum circuit.  This shows that the implementation of energy measurements on generic qudit chains is as hard as the realization of quantum computation. Here a ``measurement'' is any procedure that samples from the spectral measure induced by the observable and the state under consideration. As opposed to measurement-based quantum computation, the post-measurement state is irrelevant.}

\section{Introduction}
The characteristic feature of quantum systems is the abundance of mutually incompatible observables. According to the axioms of standard quantum mechanics \cite{Ja68} every self-adjoint operator on the system Hilbert space defines a physical variable that could in principle be measured (the richness of observables even increases if one considers generalized observables, defined by positive operator-valued measures \cite{Da76}). In particular, for systems that are composed of many components it is by no means obvious how to measure an arbitrary observable given its description as a Hilbert space operator.  

In recent years, quantum information processing has shown how one could {\it in principle} design arbitrarily complex unitary transformations by concatenating elementary operations \cite{DiV}. 
Given that a measurement procedure for one maximally abelian observable is available, such unitary transformations would allow us to measure every self-adjoint operator by transforming it into a function of the former. Moreover, it has been recognized that the so-called quantum phase-estimation procedure can be used to measure self-adjoint observables \cite{PSPACE,Trav,Habil}. Insights of this kind raise the question of which types of observables are {\it easy} to measure and which ones require {\it complex} transformations to reduce them to observables for which measurement procedures are known. Even if we specify a set of elementary observables and a set of elementary unitary transformations, it is a difficult task to find lower bounds on the required control operations. The close relation to tasks of quantum information processing suggests to address this question from a different point of view: Instead of trying to find such lower bounds (which are also hard to get for computational problems) one should rather try to place the question of the hardness of certain measurements in the context of complexity theory. This will be done in the present paper by showing that the implementation of certain measurements would already make it possible to solve classically hard computational problems.

The computational power of quantum measurements has extensively been explored in recent years since models for quantum computations have been described that are solely based on measurements
\cite{Childs-2005-71,MessRechnen,RB00}. However, these models require either concatenations of measurements \cite{MessRechnen} or the preparation of an entangled initial state \cite{RB00}. In contrast, it has been shown in \cite{PSPACE} that accurate measurements can even solve computational PSPACE-complete problems when applied to computational basis states \cite{PSPACE}. This statement requires a high measurement accuracy allowing one to resolve the exponentially small spectral gaps that are typical for interactions in many-particle systems.  Therefore, this result should be understood as exploring complexity-theoretic limitations on the measurement {\it accuracy} that can be achieved. It should not be interpreted in the sense that all (i.e., even less accurate) measurements of these observables are also difficult. 

Based on the constructions in \cite{PSPACE}, it was shown in the unpublished work \cite{WZ:06} that one obtains the computational complexity class BQP instead of PSPACE if the required measurement accuracy is only {\it inverse polynomial} instead of {\it exponential} in the number of particles. More precisely, it was proved that a measurement instrument is a full BQP-oracle if it is able to measure
so-called $k$-{\it local} observables when the system is in an appropriate basis state\footnote{Note that the computational power of measurements of ``sparse'' observables provided the underlying idea for the BQP-hardness proof of diagonal entry estimation problem in \cite{DiagonalEntryToC}}. Here and in the following the term ``basis state'' is used in the sense of a previously determined basis consisting of product states in the many-particle system. 

Here we tighten the results of \cite{WZ:06} in four respects. First, we show that $k$-locality can be replaced with finite range interactions in one-dimensional chains. Second, we can restrict the attention to translationally invariant operators. Self-adjoint operators of this kind are closer to Hamiltonians of real physical systems. Consequently, both modifications increase thus the physical relevance of the result: measurements of the observable ``energy'', i.e., the Hamiltonian $H$ of a system plays a central role in understanding its physics as the spectrum of the Hamiltonian determines the dynamical and thermodynamical behavior. Third, we do not need a measurement apparatus for a general translationally invariant Hamiltonians. Instead, it suffices in present setting to measures just one {\it specific} observable that is universal. Fourth, we need only a single-shot measurement instead of {\it repeated} measurements. 

The idea of the construction is as follows. First of all
 we construct a Hamiltonian $H$ that encodes a one-dimensional quantum cellular automaton. In analogy to the constructions in \cite{Feynman:85,Margolus:90,KitaevShen,Ergodic,ErgodicQutrits,Vollbrecht}, the time-evolution $\exp(-iHt)$ leads to superpositions between different time-steps and also between time-steps in a backward-time computation instead of implementing the computation step by step. 
Then we assume that we are given an arbitrary quantum circuit that computes the result ({\rm YES}, {\rm NO}) of a computational problem
when applied to an appropriate basis state (with respect to an a priori chosen product basis).
Since the cellular automaton is universal for quantum computing, we can chose
an initial basis state such that it simulates the quantum circuit. In other words,
the program for the computation and the classical input data are encoded into the basis state. Due to our specific construction of $H$, the spectrum  of its restriction to the smallest invariant subspace containing the initial state depends on the result of the simulated quantum circuit. 
 Since we have ensured that the two spectra corresponding to the solutions {\rm YES}  and {\rm NO} are disjoint we obtain the answer by measuring only once. 

It should be emphasized that we obtain the result of the computation by applying measurements to the {\it initial} state which is a basis state encoding program and data. The following argument describes in a metaphoric way why this is possible. We construct $H$ in such a way that the computation process stops at the end of the computation whenever the output is negative. Otherwise it continues with performing a large number of idle operations. Whether or not the computation process stops is certainly relevant for the spectrum of the restriction of $H$ to the smallest $H$-invariant subspace containing the initial state. We choose the number of idle steps in such a way that the spectra become mutually disjoint. Then we can check the result of the computation by a single-shot measurement of the energy applied to the initial state. 

\section{Defining measurements and their accuracy}\label{Meas}

Our complexity theoretic results heavily rely on the definition of what it means to measure an observables with a certain accuracy. Following standard definitions in quantum mechanics \cite{Ja68}, we define a quantum transformation to be a measurement of the observable $A$ if its application to an arbitrary state $\rho$ generates eigenvalues $\lambda_j$ of $A$ as outcomes such that ${\rm tr}(\rho Q_j)$ is the probability of obtaining $\lambda_j$, where $A=\sum_j \lambda_j Q_j$ is the spectral decomposition of $A$. In other words, an $A$-measurement (applied to the state $\rho$) allows us to sample from the {\it spectral measure} induced by the operator $A$ and the state $\rho$. Since the measurement
accuracy is relevant for the implementation complexity (compare \cite{ PSPACE}) we need a precise definition of the accuracy.

Before formally introducing approximative measurements, we point out that a process generating outcomes whose expected value coincides with ${\rm tr}(\rho A)$ is not sufficient for our purposes. The ability to implement measurement of the latter type is certainly a much weaker assumption. Consider, for instance, a Hamiltonian $H$ of the form
\[
H:=\sum_{j=1}^n H_j
\]
where each $H_j$ is an operator acting on $k$ adjacent particles only. Then we can reproduce the expected value ${\rm tr}(\rho H)=\frac{1}{n}\sum_j {\rm tr}(\rho (n\,H_j))$ by choosing a value $j$ randomly (according to the uniform distribution)  and implementing a measurement for $n\,H_j$ (requiring the ability to control only a small subsystem of the quantum system).

We now define an approximative measurement:

\begin{Definition}[Measurement accuracy]\label{Acc}${}$\\
A measurement of the observable $A$ is said to have {\it maximal error} $\delta$ and {\it reliability} $1-\epsilon$ if the following condition holds: The probability of obtaining an outcome $\lambda$ in the interval $[a-\delta,b+\delta]$ is at least ${\rm tr}(\rho Q_{[a,b]})(1-\epsilon)$ for every interval $[a,b]\subseteq\R$ and every state $\rho$, where $Q_{[a,b]}$ denotes the spectral projection of $A$ corresponding to the eigenvalues in $[a,b]$. 
\end{Definition}

We emphasize that our definition of approximate measurements focuses only on the measurement outcomes and not on the post-measurement states that are irrelevant.

It is known that the time evolution $\exp(-iAt)$ can efficiently be simulated for all $k$-local operators\footnote{Here a $k$-local operator is defined as an  operator
that can be  written as a sum of terms that act on $k$ (not necessarily adjacent) qubits only.}
 $A$ acting on $n$-qubits in the following sense: a unitary $U_t$ with $\|U_t-\exp(-iHt)\| \leq \theta$ can be implemented with resources that are only polynomial in $1/\theta$, $n$, and $t$ \cite{ChildsDiss,BACS:06,aharonov}.
For $k$-local  operators on $n$ {\it qudits} one has to introduce elementary operations on such a system (e.g. those involving only two qudits) in order to define an appropriate notion of complexity.

It was shown how to realize approximate
$A$-measurements using the quantum phase estimation procedure using approximations $V$ of $U:=\exp(-iA/\|A\|)$ for appropriate $t$ \cite{PSPACE,DiagonalEntryToC}. 
A measurement with maximal error $\delta$ and reliability $1-\epsilon$ in our sense can be achieved using only resources polynomial in $1/\delta$, $1/\epsilon$, and $n$. This  follows by adapting the arguments of \cite{PSPACE,DiagonalEntryToC,RandomWalkBQP,BioBQP} to the above definition of reliability and accuracy. 

Roughly speaking, this result means that the realization of quantum computation is at least as hard as the implementation of approximate measurements of $k$-local observables. The results of \cite{WZ:06} can be interpreted as the converse statement saying that the implementation of $k$-local measurements is at least as hard as the realization of quantum computation. To be more precise, \cite{WZ:06} showed that repeated $k$-local measurements can solve all problems in the complexity class PromiseBQP (i.e., the class of problems that can be solved by the quantum computer in polynomial time by a probabilistic algorithm) and the required measurement accuracy is polynomial in the running time of the simulated circuit. The present work shows that approximate measurements of physically more relevant Hamiltonians (translationally invariant finite-range interactions on qudit chain) can also solve all problems PromiseBQP. Before proving this, we give the formal definition of PromiseBQP.

\section{Quantum computation as approximate energy measurement}
The complexity class BQP is usually considered to represent the class of problems that can be solved efficiently on a quantum computer. However, it is often necessary to consider a generalization of this complexity class
given by the
 promise version of BQP. For example, if we want to study complete problems (i.e., problems that fully capture the power of quantum computing) then we have to work with PromiseBQP since BQP is not known to contain complete problem (just as its classical counterpart BPP and other semantic complexity classes such as MA). For these reasons, we work with PromiseBQP. 

BQP is the class of {\it language recognition} problems
that can be solved efficiently on a quantum computer.
A language recognition problem is to decide whether a given string ${\bf x}$ is an element of a language $L\subseteq\{0,1\}^*$ or not. The essential point is that  
a language defines a partition of the  set of {\it all} strings
into those that belong to $L$ and  those belonging to the complement of $L$
and that the quantum computer must be  able  to decide whether $x\in L$
and $x\not \in L$ for every $x$. 
The
promise version of a complexity class can also contain 
decision problems that
do not correspond to language recognition problems. 
Given a set of {\it allowed} inputs $\Pi:=\Pi_{\rm YES} \cup \Pi_{\rm NO}$, the  problem is to decide whether $x\in \Pi_{\rm YES}$ or $x\in \Pi_{\rm NO}$
given the promise that $x\in \Pi$ (note that language recognition problems are promise problems where $\Pi$ is the set of all strings).  
We denote this problem by $(\Pi_{\rm YES},\Pi_{\rm NO})$. 
Observe that the problem of deciding if ${\bf x}$ is in $\Pi$ could be computationally much harder.

The formal definition of PromiseBQP is:
\begin{Definition}[PromiseBQP]\label{promiseBQP}${}$\\\
PromiseBQP is the class of promise problems that can be solved by a uniform\footnote{By ``uniform circuit'' we mean that there exists a polynomial time classical algorithm that generates a sequence of a polynomial number of  
quantum gates  for every desired input length.} family of quantum circuits $(U^{(n)})$.  More precisely, it is required that this family of quantum circuits $U^{(n)}$ decides if a string $\bf{x}$ of length $n$ is a YES-instance or NO-instance in the following sense.  The application of $U^{(n)}$ to the computational basis state $|{\bf x},{\bf 0}\>$ produces the state 
\begin{equation}\label{Schalt}
U^{(n)} |{\bf x},{\bf 0}\rangle = 
\alpha_{{\bf x},0} |0\> \otimes |\psi_{{\bf x},0}\> + 
\alpha_{{\bf x},1} |1\> \otimes |\psi_{{\bf x},1}\>
\end{equation}
such that
\begin{enumerate}
\item $p_{{\bf x},1}:=|\alpha_{{\bf x},1}|^2 \geq 2/3$ for all ${\bf x}\in\Pi_{\rm YES}$ and 
\item $p_{{\bf x},0}:=|\alpha_{{\bf x},0}|^2 \geq 2/3$ for all ${\bf x}\in\Pi_{\rm NO}$ (or equivalently, that $p_{{\bf x},1}\leq 1/3$).
\end{enumerate}
Equation~(\ref{Schalt}) has to be read as follows. The input string
${\bf x}$ determines the first $n$ qubits. The remaining $a=poly(n)$
additional ancilla qubits are initialized to $|{\bf 0}\>$. After $U^{(n)}$ has been
applied we interpret the first qubit as the relevant output; the remaining $n+a-1$ output values are irrelevant. 
The number of the gates of the circuit $U^{(n)}$ is $r=poly(n)$.
\end{Definition}

Definition~\ref{promiseBQP} clarifies the notion of an ``efficient quantum algorithm''. The following theorem is the main result of the paper. It states that efficient quantum algorithms can be simulated by energy measurements whose accuracy in the sense of Definition~\ref{Acc} is inverse polynomial in the length of the simulated circuit, where the Hamiltonian is a finite-range translationally invariant Hamiltonian on a one-dimensional qudit chain and the initial state is a canonical basis state. The proof of the theorem 
follows from the construction of the Hamiltonian in Section~\ref{sec:Hamiltonian} using some general spectral analysis developed in 
the following section, phrased as
Lemma~\ref{FLemma} and Lemma~\ref{sampleLemma}.

\begin{Theorem}\label{thm:energy}
There is a family of Hamiltonians
\begin{equation}\label{eq:form}
H^{(m)} := \sum_{j=0}^{m-2} {\cal E}_j^{(m)}(V+V^\dagger)
\end{equation}
acting on $(\C^{56})^{\otimes m}$, where $V$ acts on $(\C^{56})^{\otimes 2}$ and ${\cal E}_j^{(m)}$ defines an embedding by 
\[
{\cal E}_j^{(m)}(V+V^\dagger)=\onemat^{\otimes j}\otimes (V+V^\dagger)\otimes\onemat^{m-2-j}\,,
\]
such that energy measurement of the Hamiltonians $H^{(m)}$ in canonical basis states with maximal error $\delta=1/poly(m)$ and reliability $1-\epsilon$ can probabilistically solve PromiseBQP in the following sense:

Let $U^{(n)}$ be a family of quantum circuits that solve a problem in PromiseBQP as in Definition~\ref{promiseBQP}. Then
for all input strings ${\bf x}$ of length $n$, there is a partition $\R=Y_m\dot{\cup} N_m$, and a computational basis state $|\psi_x\>\in(\C^{56})^{\otimes m}$ such that 
\begin{eqnarray}
\Pr(\lambda\in Y_m|x\in\Pi_{\rm YES}) & \ge & p_{{\bf x},1}\, (1-\epsilon) \\
\Pr(\lambda\in N_m|x\in\Pi_{\rm NO})  & \ge & p_{{\bf x},0}\, (1-\epsilon)
\end{eqnarray}
where $\lambda$ is the random variable defined by the energy measurement of $H^{(m)}$ in the state $|\psi_x\>$ and $m=poly(n)$. The quantities $m$, $|\psi_x\>$, $Y_m$, and $N_m$ can be efficiently computed from ${\bf x}$ and $U^{(n)}$.
\end{Theorem}

%
%

\section{Spectral requirements on the Hamiltonian}
\label{Sec:spectral}
We start with Hamiltonians that are not necessarily of the special form in eq.~(\ref{eq:form}). This relaxation allows us to provide a simpler and more intuitive understanding of why accurate energy measurement can solve PromiseBQP-problems at all. We construct a suitable finite-range translationally invariant Hamiltonian on a qudit chain in Section~\ref{sec:Hamiltonian}.

We now briefly review the basic principles of the autonomous Hamiltonian computers proposed by {\sc Benioff}, {\sc Feynman}, and {\sc Margolus} \cite{Benioff,Feynman:85,Margolus:90} since our Hamiltonian can be view as a special type of an autonomous Hamiltonian quantum computer. Originally, the construction of computational models in terms of autonomous Hamiltonian dynamics of closed physical systems was motivated,  among others, by the study of thermodynamics of computation. 
{\sc Kitaev} observed that these ideas are also useful in quantum complexity theory since they provide a connection between problems of determining spectral properties of Hamiltonians and those of finding solutions of hard computational problems. This link has mainly been used in the context of the proof the QMA-completeness of the so-called local Hamiltonian problem (estimating the ground state energy for a wide range of Hamiltonians \cite{KitaevShen,KempeRegev,Kempe2local,Oliveira,Gottesman}). 

Let $U=U_r U_{r-1} \cdots U_1$ be a quantum circuit consisting of the elementary gates $U_t$ acting on the logical Hilbert space $\cH$ for $t=1,\dots,r$.
We adjoin a Hilbert space $\cH_{clock}$ that represents the clocking device of the autonomous computer and define on $\cH\otimes \cH_{clock}$
the ``forward time operator'' 
\begin{equation}\label{F}
F:=\sum_{t\in \Z} U_t \otimes |t+1\rangle \langle t|\,,
\end{equation}
where we have used the convention $U_t={\bf  1}$ for $t\not\in \{1,\dots,r\}$.
The states $|t\>$ label some states in $\cH_{clock}$ and represent the time steps of the quantum circuit $U$. 
They do not necessarily span the whole space $\cH_{clock}$. 
We then define a Hamiltonian by
\[
H := F + F^\dagger\,.
\]
Applying the
time evolution $\exp(-iHt)$ to the state $|{\bf x},{\bf 0}\rangle \otimes |0\rangle$ leads to superpositions
of states of the form 
\[
F^t |\psi\> = \big( \prod_{j=1}^t U_j |{\bf x},{\bf 0}\> \big) \otimes |t+1\> \quad \hbox{ with } \quad t>0\,,
\]
and
\[
|{\bf x},{\bf 0}\rangle \otimes |t\rangle \hbox{ with } t\leq  0\,.
\]
The idea of Hamiltonian computers  is then to initially prepare the clock in a superpsoition  of states with negative $t$ in such a
way that the wave packet propagates mainly in forward direction and triggers the implementation of gates.

The reason why such models of autonomous computation can only simulate quantum circuits in such a ``broader sense'' (i.e.,
one has  always superpositions of different computational states) is that the Hamiltonians should 
be $k$-local 
for some small constant $k$. It is not known how to directly implement unitary operations $U$ by time-independent Hamiltonians $H$ of this type where $U=\exp(-iH)$ represents some interesting computation. 

An important feature of the forward-time operators in the above mentioned literature is that they act unitarily on the relevant subspace of ``computational states'' spanned by $F^t (|{\bf x},{\bf 0}\>\otimes |1\>)$  and $(F^\dagger)^t (|{\bf x},{\bf 0}\>\otimes |1\>)$
with  $t\in \N$.  
This makes a spectral analysis feasible since $F$ and $F^\dagger$ commute on this subspace. 

For our construction, we do  not want the Hamiltonian really 
to perform the computation in the above sense.  However, the above  explanation still remains
the leading intuition. 
To prove that approximate energy measurements can solve PromiseBQP-problems, we have to introduce a modification of the above forward-time operator such that it is no longer unitary on the relevant subspace and analyze its spectral properties. We need the {\it read-out} gate $R:=|1\>\<1|$ and the {\it annihilation} gate $A:={\bf 0}$ (where ${\bf 0}$ is the zero matrix) for this modification. Let $U=U_r U_{r-1} \ldots U_1$ be a quantum circuit as above and let $s$ be greater than $r$. We construct a non-unitary quantum circuit
\[
\hat{U}=\hat{U}_{s+1}\hat{U}_s \hat{U}_{s-1} \cdots \hat{U}_{r+1} \hat{U}_r \hat{U}_{r-1} \cdots \hat{U}_1 \hat{U}_0
\]
where its elementary gates are defined as follows
\begin{eqnarray}
\hat{U}_0     & := & A  \label{Uhatdef1}\\
\hat{U}_t     & := & U_t \mbox{ for $t=1,\ldots,r$} \\
\hat{U}_{r+1} & := & R \\
\hat{U}_t     & := & I \mbox{ for $t=r+2,\ldots,s$} \\
\hat{U}_{s+1} & := & A \label{Uhatdef5}\,.
\end{eqnarray}
The reader  may be confused that $\hat{U}$ is just the trivial operator  ${\bf 0}$. The reason why we define such an unusual ``quantum circuit''  
is that this definition allows us to interpret every operation performed by the forward time operator, including
state annihilation, as a (not necessarily unitary) quantum ``gate''.
Then we  construct a modified forward time operator  (compare eq.~(\ref{F})) by
\[
F:=\sum_{t=0}^{s+1} \hat{U}_t  \otimes |t+1\rangle \langle t|\,.
\]

To intuitively understand why the spectral  measure of the Hamiltonian $H:=F+F^\dagger$ will then reflect the solution of the
computational problem we observe the following. Assume we apply the time  evolution $\exp(-iHt)$ to the state 
$|{\bf x},{\bf 0}\> \otimes |1\>$  and the answer  of the problem is ``YES'' with probability $1$.  In this case 
the orbit 
\[
\Big(\exp(-iHt)(|{\bf x},{\bf 0}\> \otimes |1\>)\Big)_{t\in \R_+}
\]
 consists of superpositions of the states
$F^t(|{\bf x},{\bf 0}\> \otimes |1\>)$ for $t=0,\dots,s$. This is because $F^\dagger$ and $F$ 
annihilate the first and the last state in this set, respectively, and on the remaining states (i.e. $t=0,\dots,s-1$) 
they commute since the non-unitary readout gate is irrelevant. The dynamics is then mathematically equivalent to a particle moving on a chain of length $s+1$ (with a dynamics induced by
``hopping terms''). 

If the answer is ``NO'' with probability $1$ the dynamics is restricted to the subspace spanned by
$F^t(|{\bf x},{\bf 0}\> \otimes |1\>)$ for $t=0,\dots,r$, i.e., a dynamics on  a chain of length $r+1$. 

In the generic case the answer is non-deterministic and we obtain a mixture of both  cases such
that probability distribution of the  results of  a $H$-measurement depends directly on the
outcome probabilities of  the  simulated circuit. For the mathematical analysis of this case 
we will use the decomposition
\[
|{\bf x},{\bf 0}\>\otimes |1\> = \sqrt{1-p_1}|\psi_0\> \oplus \sqrt{p_1}|\psi_1\>
\]
where $|\psi_1\>$ is obtained by renormalizing $(F^\dagger)^{r+1} F^{r+1} (|{\bf x},{\bf 0}\>\otimes |1\>)$ and
$|\psi_0\>$ by renormalizing $|{\bf x},{\bf 0}\>\otimes |1\> - (F^\dagger)^{r+1} F^{r+1} (|{\bf x},{\bf 0}\>\otimes |1\>)$.
These components correspond to new initial states for which the answer in the readout step is deterministic.  
The  following  lemma specifies precisely the statement and the conditions on $F$ that we require.

\begin{Lemma}\label{FLemma}
Let $U=U_r U_{r-1} \cdots U_1$ be a quantum circuit that accepts the state $|{\bf x},{\bf 0}\>$ with probability $p_{{\bf x},1}$ in  the
sense of Definition~\ref{promiseBQP}.
Let $\hat{U}=\hat{U}_{s+1} \hat{U}_s \cdots \hat{U_0}$ be the corresponding non-unitary quantum circuit as defined in eqs.~(\ref{Uhatdef1}) to (\ref{Uhatdef5}).
Let $F$ be an operator acting on $\cH_d\otimes \cH_{clock}$ where $\cH_d$ stands for ``data space'' and $\cH_{clock}$ for ``clock space'',
respectively. Let $|0\>,\dots,|s+1\>$ denote some mutually orthogonal states of the clock register.  Assume that $F$ and $F^\dagger$ act as 
\begin{eqnarray}
F(|\phi\> \otimes |t\>)           & = & \hat{U}_t|\phi\> \otimes |t+1\> \\
F^\dagger(|\phi\> \otimes |t+1\>) & = & \hat{U}^\dagger_t|\phi\> \otimes |t\>\,,
\end{eqnarray}
for $t=0,\ldots,s$ on the smallest $F$ and $F^\dagger$ invariant subspace containing the initial state $|\psi\>:=|{\bf x},{\bf 0}\> \otimes |1\>$.

Then the spectral measure induced by $F+F^\dagger$ and the initial state $|\psi\>$ is given by the convex combination
\[
(1 - p_{{\bf x},1}) \cP_{r+1} + p_{{\bf x},1} \cP_{s+1}
\]
where $P_\ell$ denotes the spectral measure induced by the adjacency matrix of the line graph with $\ell$ vertices (for $\ell=r+1$ and $\ell=s+1$)
and the first basis vector $|0\>$. To be more explicit,
the adjacency matrix is given by
\[
\cL_\ell:=S_\ell+S_\ell^\dagger \,,
\]
where $S_\ell$ denotes the non-cyclic shift in $\ell$ dimensions with the canonical basis vectors $|0\>,|1\>,\ldots,|\ell-1\>$, i.e.,
\[
S_\ell =\sum_{v=0}^{\ell-2} |v+1\>\<v|\,.
\]
Let $\lambda_v$ be the $v$th eigenvalue of $\cL_\ell$ and $Q^{(\ell)}_v$ be the corresponding spectral projection. 
Then $P_\ell$ is the probability distribution on the eigenvalues defined by 
\[
P_\ell(\lambda_v) := \<0| Q^{(\ell)}_v |0\>\,.
\]
\end{Lemma}

\noindent
{\bf Proof:} Let $|\psi_1\>$ be the vector obtained by renormalizing the vector
\[
(F^\dagger)^{r+1} F^{r+1} |\psi\>=(U_1^\dagger\cdots U_r^\dagger R U_r\cdots U_1|{\bf x},{\bf 0}\rangle)\otimes |1\rangle 
\]
 and $|\psi_0\>$ by renormalizing $|\psi\> - (F^\dagger)^{r+1} F^{r+1} |\psi\>$. Let $\cV_0$  be  the span  of $\{F^t|\psi_0\>\}_{j=0,\dots,r}$ and $\cV_1$ the span of $\{F^t|\psi_1\>\}_{j=0,\dots,s}$. The statement $\cV_0\perp  \cV_1$ follows from the fact that
\begin{equation}\label{FtFt}
F^t|\psi_0\rangle \perp F^{t'} |\psi_1\rangle
\end{equation}
holds for all $t\neq t'$, 
because the states then correspond to different basis vectors in  the  clock space. 
For  $t=t'=r$ 
the  restrictions of the  states $F^t|\psi_1\rangle$ and $F^t|\psi_0\rangle$ to  the
data register are given by renormalizing  
\begin{equation}\label{cpF}
R U_r \cdots U_1|\psi\rangle
 \quad \hbox{ and } \quad ({\bf1} -R) U_r \cdots U_1|\psi\rangle\,,
\end{equation}
respectively. For $0\leq t\leq r-1$ we have to apply the unitary operator  
$U^\dagger_{t+1} \cdots U_r^\dagger$ to both states in eq.~(\ref{cpF}), which obviously preserves the orthogonality relation.  

One easily checks that $\cV_0$ and $\cV_1$ are invariant under the action of $F$ and $F^\dagger$.
This implies that they are also invariant under $F+F^\dagger$. 

We first determine the spectral measures induced by $|\psi_0\>$ and $|\psi_1\>$. On $\cV_0$ the operator $F$ is unitarily equivalent to the non-cyclic shift $S_{r+1}$ by identifying 
the orthonormal basis vectors $F^t|\psi_0\>$ with the basis vectors $|t\> \in \C^{r+1}$ for $t=0,\dots,r$. This unitary equivalence is readily verified by checking that $F$ maps the $t$th  basis vector to the $(t+1)$th for $t<r$ and annihilates $F^r|\psi_0\>$. Hence the spectral measure induced by $F+F^\dagger$ on $|\psi_0\>$ is the same as the spectral measure induced by $\cL_r=S_{r+1}+S_{r+1}^\dagger$ on $|0\>$, i.e., it is equal to $\cP_{r+1}$.  

Similarly,  $F+F^\dagger$ induces the spectral measure $\cP_{s+1}$ on $|\psi_1\>$. The spectral measure induced by $|\psi\>=\sqrt{1-p_{{\bf x},1}}|\psi_0\> \oplus \sqrt{p_{{\bf x},1}} |\psi_1\>$ is then just the convex sum of the measures induced by $|\psi_0\>$ and $|\psi_1\>$, respectively.  By elementary linear algebra, the corresponding weights are $1-p_{{\bf x},1}$ and $p_{{\bf x},1}$. This completes the proof. \hfill $\square$

\vspace{0.3cm}
\noindent
The lemma shows how to construct a Hamiltonian $H=F+F^\dagger$ such that the distribution of 
the results in an {\it accurate} measurement is given by the measure
$(1-p_{{\bf x},1}) \cP_{r+1} + p_{{\bf x},1} \cP_{s+1}$ as in Lemma~\ref{FLemma}. To analyze how to assign measurement outcomes with ``YES'' and ``NO''
we recall that the spectrum of the line graph $\cL_\ell$ of length $\ell$ is given by the values 
\[
\lambda_v=2\cos\left( \frac{\pi (v+1)}{\ell+1} \right)
\]
for $v=0,\dots,\ell-1$ \cite{Cvetkovic}. Let $\cS_{r+1}:=\{\lambda_v\,:\,v=0\ldots,r\}$ be the spectrum of the line graph $\cL_{r+1}$ and $\cS_{s+1}:=\{\mu_w\,:\, w=0,\ldots,s\}$ that of $\cL_{s+1}$.

If $r+2$ and $s+2$ are relatively prime, then $\cS_{s+1}$ and $\cS_{r+1}$ are disjoint. This assumption is useful
since it will allow us to replace the quantum computation with a {\it single shot} measurement instead of {\it repeated} sampling. 
To obtain a lower bound on the distance between these two sets we
apply the function $x \mapsto \arccos (x/2)/\pi$ to the eigenvalues $\lambda_v$ and $\mu_w$ and obtain the values
\[
a_v:=\frac{v+1}{r+2} \quad \hbox{ and } \quad b_w:=\frac{w+1}{s+2}\,.
\]
The minimal distance between any of these values is at least 
$1/(r+2)(s+2)$. 
On the interval $[0,\pi]$ the cosine function satisfies the inequality 
\[
|\cos (x)- \cos (y)|\geq \frac{1}{4} (x-y)^2\,,
\]
for all $x,y$ with  $|x-y|\leq 1$. 
These facts imply that the minimal distance between the spectra $\cS_{r+1}$ and $\cS_{s+1}$ is at least
\begin{equation}\label{eq:spectralDistance}
\Delta:= \frac{\pi^2}{2(r+2)^2(s+2)^2}\,.
\end{equation}
From an intuitive point of view, this already shows that the ability to implement measurements with inverse polynomial error 
makes it possible to assign a measurement outcome to either of the two spectra.  The following lemma bases this statement on the 
formal definition of measurement accuracy and analyzes the probability of correctly classifying the input. It follows immediately from the arguments above.

\begin{Lemma}\label{sampleLemma}
Let the distribution of measurement results  in an accurate measurement be  given by 
\[
(1-q) \cP_{r+1} + q \cP_{s+1}
\]
with $q\in  [0,1]$, where $\cP_{\ell}$ for $\ell=r+1,s+1$ denotes the spectral measure
of the line graphs of  length  $\ell$ as above. 
Assume we have an approximative 
measurement procedure with accuracy $\Delta$ as in eq.~(\ref{eq:spectralDistance}) and reliability $1-\epsilon$.
For $\delta:=\Delta/3$ define
\[
Y:=\bigcup_{w=0}^s \, [\mu_w - \delta, \mu_w + \delta] \mbox{ and } N:=\R\setminus Y\,,
\]
where $\mu_w$ denotes  the $w$th eigenvalue of the  line graph of length  $s+1$.  
Let $\lambda$ be the measurement outcome.

Then we have the conditional probabilities 
\begin{eqnarray}
\Pr(\lambda\in Y\, | \, {\bf x}\in\Pi_{\rm YES}) & \ge & q\, (1-\epsilon) \\
\Pr(\lambda\in N\, | \, {\bf x}\in\Pi_{\rm NO})  & \ge &  (1-q) \, (1-\epsilon)\,.
\end{eqnarray}
\end{Lemma}

The  lemma
shows that the error of the measurement-based  decision procedure then essentially reproduces the error probabilities
of the simulated circuit provided that the probability to have  an error  greater than $\Delta$ is small.   
The arguments in this section thus establish that energy measurements for local Hamiltonians  
can be used to solve PromiseBQP-problems.  Here ``local'' means  that the  interaction involves only a few, but  not necessarily 
adjacent, qubits (if we represent the clock also by a qubit register).  
Unfortunately, this does not suffice to prove Theorem~\ref{thm:energy} because we have not described how 
to construct a {\it nearest-neighbor translationally invariant interaction}  $H$ on a qudit chain having the desired properties. In the next section, we show how to construct such interaction based on a quantum cellular automaton (QCA) in such a way that it still satisfies the spectral properties in Lemma~\ref{FLemma}. Our interaction is not supposed to be the simplest operator satisfying all the above requirements. However, we have chosen an encoding of quantum circuits that seems to be rather concise and quite natural. It is likely that more sophisticated encodings could significantly reduce the size of the unit cell of the QCA.

\section{Construction of the Hamiltonian}\label{sec:Hamiltonian}

To construct the QCA, we restrict ourselves to the following universal set of quantum gates. Let $S$ denote the swap gate and $W$ the controlled gate defined by
\[
W:=\left(\begin{array}{cccc} 1 & 0 & 0 & 0 \\ 0 & 1 & 0 & 0 \\ 0 & 0& \frac{1}{\sqrt{2}} & -\frac{1}{\sqrt{2}} \\ 0 & 0 & \frac{1}{\sqrt{2}} & \frac{1}{\sqrt{2}} \end{array} \right)\,.
\]
We require that both gates be only applied to adjacent qubits. Moreover, we require that $W$ be applied only in one direction, i.e.,
the control-wire will  always be  the  left qubit of the corresponding pair. 
 This set of gates is universal because the $W$ gate is universal if it can be applied to arbitrary qubit pairs \cite{Shepherd}. We assume without loss of generality that the quantum circuits $U$ in Definition~\ref{promiseBQP} are composed only of $S$ and $W$ acting on adjacent qubits. We also assume that gates acting on disjoint qubit pairs can be performed in parallel. In this way, the quantum circuits are split into {\it layers}. We refer to these layers as the {\it time steps} of the quantum circuit.

\subsection{Description of our quantum cellular automaton}

We now construct a quantum cellular automaton (QCA) that can simulate any such quantum circuit such that the total number of time steps required is
polynomial in the number of gates and the number of qubits used in the  simulated circuit.  

We encode the quantum circuit and its input in the initial state of the QCA. It acts on
\[
\cH:=\cH_c^{\otimes m}
\]
where $\cH_c:=\C^{56}$ denotes the Hilbert space of a single cell. Our Hamiltonian has the form
\begin{equation}\label{eq:ourHamiltonian}
H=F + F^\dagger = \sum_{j=0}^{m-2} {\cal E}^{(m)}_j(V + V^\dagger )\,,
\end{equation}
where $V$ acts on two cells, i.e., on $\cH_c \otimes \cH_c$. Each cell consists of a data and a program cell, i.e., we have
\[
\cH_c:=\cH_p \otimes \cH_d
\]
where $\cH_p$ and $\cH_d$ denote the Hilbert space of the program and data cell, respectively. We refer to the set of all program cells (i.e., $\cH_p^{\otimes m}$) as the {\it program band} and to that of all data cells (i.e., $\cH_d^{\otimes m}$) as the {\it data band}. The program band initially contains the description of the quantum circuit $U$ to be simulated.  We refer to this description as the {\it program code}. This code is divided into {\it blocks} where each block corresponds to one time step of $U$. The {\it data register} is a subset of the data band that corresponds to the register on which $U$ acts, i.e., it represents the logical qubits. The remaining part of the data band will only contain
formatting symbols. 

The Hilbert spaces of the data and program cells are denoted by $\cH_d=\C^{4}$ and $\cH_p=\C^{14}$, respectively. The basis vectors of $\cH_d$ are identified with the symbols in
\[
\{0, 1\} \cup \{\|, \bullet \}\,.
\]
those of of $\cH_p$ with the symbols in
\[
\{\sI, \sS, \sW, \sR, \sA\} \cup \{ \mI, \mS, \mW, \mR, \mA\} \cup \{\square,\blacksquare, \Diamond, \#,\}\,.
\]
The program code is composed of the {\em gate symbols} $\sI$, $\sS$, $\sW$, $\sR$, and $\sA$. These symbols correspond to the identity, $S$, $W$, $R$, and $A$, respectively.
The {\em marked gate symbols} $\mI$, $\mS$, $\mW$, $\mR$, and $\mA$ are used to control the propagation of the program code. The symbols $\square$, $\blacksquare$, $\Diamond$, and $\#$ are also used for this purpose.  We refer to these symbols as the {\em hole}, {\em execution}, {\em turn-around}, and {\em blank} symbols, respectively.  The kets $|0\>$ and $|1\>$ correspond to the two orthogonal states of a qubit. The symbols $\|$ and $\bullet$ are used for formatting purposes; the symbol $\|$ makes it possible to determine when the program has been moved by one block. They are contained only in those data cells
that do not belong to the {\it data register}.

During the dynamics the content of the program cells slides over the data cells and triggers the implementation of the gates on the data register. Since we want to construct a nearest neighbor interaction the sliding of the program can only be realized by moving the content cell by cell. Therefore, we need a special control mechanism that ensures that the program is only executed after it has been moved by exactly one block, i.e., the block is again aligned with the corresponding data cells on which the gates are supposed to act. The forward time operator $F$ in eq.~(\ref{eq:ourHamiltonian}) must implement the propagation and the execution of the gates whenever the program is aligned with the data band. 

The operator $V$ in eq.~(\ref{eq:ourHamiltonian}) is thus the product 
\[
V=T \, X\,,
\]
where $X$ applies gates on $\cH_d\otimes\cH_d$ conditioned on the state of $\cH_p\otimes\cH_p$ and 
$T$ realizes transitions between basis states of $\cH_p\otimes\cH_p$. 

Before we define $T$ and $X$ explicitly we look at a simple example to see how quantum circuits and inputs are encoded in the initial states. We also explain how the execution of the program code is controlled in a purely local procedure where no global clocking is available.

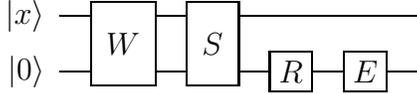
\begin{figure}
\hspace{2cm}
\Qcircuit @C=1em @R=.7em {
\lstick{|x\>} & \multigate{1}{W} & \multigate{1}{S} & \qw      & \qw      & \qw \\
\lstick{|0\>} & \ghost{W}        & \ghost{S}        & \gate{R} & \gate{E} & \qw
}
\caption{\label{fig}Example of a simple quantum circuit that acts on $2$ qubits and consists of $5$ time steps. The first qubit takes the input is $x$. The second qubit is an ancilla qubit.}
\end{figure}

\begin{enumerate}
\item Assume we want to simulate the circuit in fig.~\ref{fig}. Then we initialize the program and data bands as shown in fig.~\ref{init}. The program code is $\sA\blacksquare\sW \, \sI\sI\sS\ \, \sI\sI\sR \, \sI\sI\sA\ \, \sI$, where the gaps are used to indicate the different time steps of the quantum circuit $U$. All other program cells contain the blank symbol $\#$.
Note that the two qubit gates $W$ and $S$ are encoded by symbol pairs $\sI\W$ and $\sI\sS$, respectively, and that the program
contains an extra symbol $\sI$ 
between the blocks (i.e., the layers of the circuit) and the first command of the program reads $\sA$. 
In  the  initial state, 
the first layer of the circuit is aligned with the data register that is initialized in the state $|x,0\>$.  The symbols $\|$ to the left and right of the data register enclose exactly two $\bullet$ symbols, corresponding to the number of qubits of $U$. 

\begin{figure}
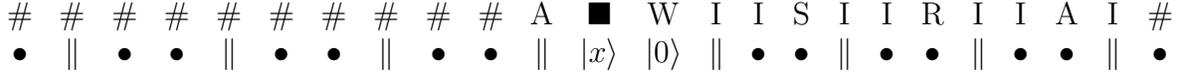

\[
\begin{array}{cccccccccccccccccccccccccccccc}
\# &\# & \# & \# & \# & \# & \# & \# & \# & \# & \sA & \blacksquare & \sW & \sI & \sI & \sS & \sI & \sI & \sR & \sI & \sI & \sA & \sI & \# \\
\bullet &\| & \bullet & \bullet & \| & \bullet & \bullet & \| & \bullet & \bullet & \| & |x\> & |0\> & \| & \bullet & \bullet & \| & \bullet & \bullet & \| & \bullet & \bullet & \| & \bullet 
\end{array}
\]
\caption{\label{init}Initialization of the program and data bands}
\end{figure}

\item It is important that the program code starts with the pair $\sA\blacksquare$. One may rather  expect the initial configuration
$\blacksquare \sA$, but this would lead to an execution of $\sA$ in the first step. 
This is analogous to Section~\ref{Sec:spectral}, where we have to start in the clock state $|1\>$ instead of $|0\>$ to prevent the execution of the first annihilation gate $\hat{U}_0:=A$. The purpose of this choice of the initial state is to ensure that it is annihilated by 
the operator $F^\dagger$ because it implements the program  in backward direction.
The execution symbol $\blacksquare$ is propagated cell by cell to the end of the program code until it arrives at the rightmost cell not containing the blank symbol $\#$. In each step, $\blacksquare$ and the gate symbol $\sG$ on its right side swap their positions. More precisely, if $\blacksquare$ is in the $j$th program cell and $\sG$ in the $(j+1)$th cell, then $\sG$ moves to the $j$th cell and $\blacksquare$ to the $(j+1)$th cell. While swapping the symbols, the gate corresponding to $\sG$ is executed.  It is applied to data cells $j$ and $j+1$ if it is a two qubit gate and to data cell $j+1$ if it is a single qubit gate. This only happens only if the data cell $j$ {\em and} data $j+1$ are inside the data register in the case of a two qubit gate. Similarly, it only happens if the data cell $j+1$ is inside the data register in the case of a singe qubit gate.

\item Once the execution symbol $\blacksquare$ has passed the end of the program code it is converted to the blank symbol $\#$ (via the creation of the intermediate turn-around symbol $\Diamond$) and a signal is sent to the begin of the program code. This signal indicates that the execution of the first time step of the program code has been completed.  To avoid that the propagation of this signal leads to a backward propagation of the program code the signal cannot occupy a program cell on its own. It propagates by converting each gate symbol $\sG$ into a marked version $\mG$ cell by cell. Once the marked gate symbol $\bG$ is at the begin of the program code it is converted into the hole symbol symbol $\square$ via the creation of the turn-around symbol $\Diamond$.

\item The hole symbol $\square$ propagates cell by cell to the end of the program where it is converted to the blank symbol $\#$ and triggers the left propagating marker $\bG$ (via the creation of the  intermediate turn-around symbol $\Diamond$).  The arrival of this marker at the begin of the program code triggers the conversion of the next copy of $\#$ into $\square$. This procedure is repeated until the begin of the program code is again aligned with the next copy of the format symbol $\|$ in the data band.  In this case, the marker $\bG$ triggers the creation of $\blacksquare$ instead of $\square$ and the whole cycle starting in the second step is repeated (again, this is done by the creation of the intermediate turn-around symbol $\Diamond$).
\end{enumerate}

The above procedure is implemented by the following nearest-neighbor transition rules. These rules depend only on the contents of two adjacent cells of the QCA. We use $*$ to denote any of the symbols $0$, $1$, and $\bullet$. The symbol represented by $*$ is left unchanged by the corresponding transition rule.  In the transition rules 2 and 6 the left lower and right lower corners are left empty to indicate that the symbol at that place is not important for the transition rule and that it is left unchanged.

\medskip
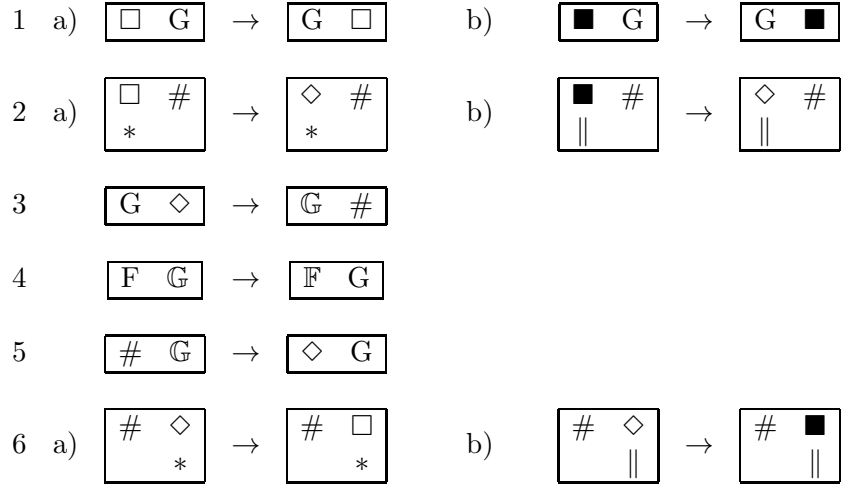
\[
\begin{array}{ccccccccccc}
1 & \mbox{a)} & \band{\square}{\G}  & \lra & \band{\G}{\square}   & & & \mbox{b)} \quad &\,\,\band{\blacksquare}{\G} & \lra & \band{\G}{\blacksquare} \\ \\
2 & \mbox{a)} & \doubleband{\square}{\#}{*}{}& \lra & \doubleband{\Diamond}{\#}{*}{} & & & \mbox{b)} \quad &\,\,\doubleband{\blacksquare}{\#}{\|}{} & \lra & \doubleband{\Diamond}{\#}{\|}{} \\ \\
3 &           & \band{\G}{\Diamond} & \lra & \band{\bG}{\#} \\ \\
4 &           & \band{\F}{\bG}      & \lra & \band{\bF}{\G} \\ \\
5 &           & \band{\#}{\bG}      & \lra & \band{\Diamond}{\G} \\ \\
6 & \mbox{a)} & \doubleband{\#}{\Diamond}{}{*}  & \lra & \doubleband{\#}{\square}{}{*} && & \mbox{b)} \quad &\,\,\doubleband{\#}{\Diamond}{}{\|} & \lra & \doubleband{\#}{\blacksquare}{}{\|} 
\end{array}
\]

\medskip
Transitions 1 a) and b) implement the rightward propagation of the symbols $\square$ and $\blacksquare$, respectively. Transitions 2 a) and b) takes place when $\square$ and $\blacksquare$ have passed the end of the program code, respectively.  They create the symbol $\Diamond$ (turn-around). Transition 3 creates a marked gate symbol that initiates the leftward moving signal.  Transition 4 implements the propagation of this signal.  Once this signal has arrived at the begin of the program code, transition 5 and transitions 6 a) and b) create the symbols $\square$ and $\blacksquare$ via the creation of the intermediate turn-around symbol $\Diamond$, respectively. The execution symbol $\blacksquare$ is created only if the turn-around symbol $\Diamond$ is exactly above the formatting symbol $\|$, which happens only if the blocks of the program code are aligned with the data register.  Otherwise, the hole symbol $\square$ is created.

Now we are ready to define the transition operator $T$. It is the annihilation-creation operator on $\cH_c\otimes\cH_c$ defined by all the above transition rules. It annihilates all configurations that do not appear on the left-hand side of any of the above transitions.

We denote by $X$ the operator that realizes the controlled execution of the gates. It is defined by
\[
\begin{array}{cccccccccccc}
X & := & |\blacksquare \sS\>\<\blacksquare \sS| & \otimes & S_d              & + & |\blacksquare \sW\>\<\blacksquare \sW| & \otimes & W_d                & + \\
  &    & |\blacksquare \sR\>\<\blacksquare \sR| & \otimes & (I_d\otimes R_d) & + & |\blacksquare \sA\>\<\blacksquare \sA| & \otimes & (I_d\otimes E_d)\, & + \\
  &    & Q & \otimes & (I_d\otimes I_d)
\end{array}
\]
where 
\[
Q:=I_p\otimes I_p \,\, - \!\! \sum_{\sG\in\{\sS,\sW,\sR,\sA\}} |\blacksquare\sG\>\<\blacksquare\sG|
\]
The projectors $|\blacksquare\sG\>\<\blacksquare\sG|$ for $\sG\in\{\sS,\sW,\sR,\sA\}$ act on $\cH_p\otimes\cH_p$. $I_p$ and $I_d$ act as identity on $\cH_p$ and $\cH_d$, respectively. $S_d$ and $W_d$ act as $S$ and $W$, respectively, on the subspace of $\cH_d\otimes\cH_d$ spanned by $\{|00\>,|01\>,|10\>,|11\>\}$. Similarly, $R_d$ and $E_d$ act as $R$ and $E$, respectively, on the subspace of $\cH_d$ spanned by $\{|0\>,|1\>\}$.

\subsection{Spectral properties}

We denote the initial state of the QCA by $|{\bf x},{\bf 0}\> \otimes |1\>$ where $|\bf{ x},{\bf 0}\rangle$ is the initial state of the quantum circuit to be simulated and $|1\rangle$ represents the state of the program band and the state of all data cells outside the data register that  corresponds to an appropriately chosen initial configuration. 

The application of $F$ changes the configuration in the program cells into a new basis state
and applies, at the same time,  gates to the relevant section in the data cells.
This is because the transition rules are designed such that only one transition is possible in each step provided that
the initial configuration has been chosen as described in the preceding section. We can thus denote the configurations as 
the clock states   $|1\rangle,|2\rangle,\dots$. 
Since our simulation of the original   circuit contains not only {\it execution}  steps but also operations
where only signals are propagated the circuit is thus extended by  polynomial overhead of identity gates
on the data cells. We thus have  $\tilde{s}+1$ steps instead  of $s+1$  until the state is completely annihilated.
Likewise, the readout  gate causes a conditional annihilation in the step $\tilde{r}+1$ instead of $r+1$.   

To make sure that $F$ and the initial state $|{\bf x,0}\rangle \otimes |1\rangle$ satisfy
the  conditions of Lemma~\ref{FLemma}  we have to 
check that  
the  application of $F^\dagger$ to one of the states in the orbit $F^t(|{\bf x,0}\rangle \otimes |1\rangle)$ with
$t=1,\dots,\tilde{s}$ indeed leads to the preceding clock state and to an annihilation for $t=0$. 
The annihilation property 
is satisfied since we have started the system in a state where the running of the program in backward  direction would
immediately execute the $\sA$ gate.
To see that $F^\dagger$ leads in the general case 
to the preceding  states, we have  to check our list of transitions in {\it backward} direction.

Every configuration that arises from the repeated application of the forward-time operator $F$ always contains exactly one element in the set $\{\square,\blacksquare,\Diamond, \mI, \mS, \mW\}$.
If the configuration contains one of the symbols $\square$ and $\blacksquare$ but they are neither preceded nor followed by the blank symbol $\#$ (i.e., they are not located at the boundaries of the program code), then only rule 1 a) or b) can be active in the backward direction.  The pattern in rule $2$ only occurs when the symbols $\square$ or $\blacksquare$ are located at the right boundary of the program code.  In this case, no other rules are active in the backward direction.  Similarly, the pattern in rule $6$ only occurs when the symbols $\square$ or $\blacksquare$ are located at the left boundary of the program (i.e., only when they are preceded by $\#$).  In this case, no other rule is active in the backward direction.

The symbol $\Diamond$ only appears in rules $2$ and $5$ on the right hand side of the transitions.  But there is no configuration where both rules can be applied backwards at the same time. Likewise, the marked gate symbols only occur in rules $3$ and $4$.  Again, there is no configuration where both rules can be applied backwards at the same time. 

Hence we have shown that $F$ meets the requirements of Lemma~\ref{FLemma} and that the spectral measure induced by $F+F^\dagger$ on the initial state $|{\bf x},0\rangle \otimes |1\rangle$ 
depends on the acceptance probability $p_{{\bf x},1}$. To design a program code such that the supports of the two corresponding spectral measures to be distinguished are disjoint we may choose the position of the annihilation gate appropriately.  In this way we can always achieve that $\tilde{r}+2$ and $\tilde{s}+2$ are relatively prime, ensuring that the minimal distance between the supports of the spectra satisfies eq.~(\ref{eq:spectralDistance}).

\section{Conclusions}
In contrast to usual measurement-based approaches to quantum computing we have constructed an observable
whose measurements have full  quantum computation power when applied to a {\it basis} state only {\it once}.
The required accuracy scales inverse polynomially with the number of gates of the simulated quantum circuit.

The observable is a Hamiltonian of a finite-range interaction with translation symmetry. Even though our specific construction
is unlikely to be found in real physical systems our result strongly suggests that energy measurements for real physical systems
is a task whose realization is as challenging as the realization of quantum computing. 
This shows, once more, that quantum information processing in a broader sense is not only required to solve {\it  computational} problems. Instead, it is a task that occurs already in the context of usual quantum control.  Since we have shown in a previous paper that 
more accurate measurements of observables of a similar type solve all problems in the complexity class PSPACE our result
can also be interpreted as showing how the complexity of quantum control procedures depend on the demanded accuracy. 

\section*{Acknowledgment}
P.W. acknowledges the support by NSF grant CCF-0726771.  S.Z. is supported by NSF grant PHY-0456720 and ARO grant W911NF-05-1-0294. We would like to thank Daniel Nagaj for very helpful discussions about the transition rules of the quantum cellular automaton.


\begin{thebibliography}{10}

\bibitem{Gottesman}
D.~Aharonov, D.~Gottesman, and J.~Kempe, 
The power of quantum systems on a line,
arXiv:quant-ph/0705.4077, 2007.

\bibitem{aharonov}
D.~Aharonov and A. Ta-Shma.
Adiabatic quantum state generation and statistical zero knowledge. 
{\em In Proc. 35th Annual ACM Symp. on Theory of Computing}, pp. 20-–29, 2003.

\bibitem{Benioff}
P.~Benioff.
\newblock {The computer as a physical system: A microscopic quantum mechanical
  model of computers as represented by Turing machines}.
\newblock {\em J. Stat. Phys.}, 22(5):562--591, 1980.

\bibitem{BACS:06}
{D. W.} Berry, G.~Ahokas, R.~Cleve, and {B. C.} Sanders.
\newblock Efficient quantum algorithms for simulating sparse {Hamiltonians}.
\newblock {\em Comm. Math. Phys.}, 270(2):359--371, 2007.

\bibitem{ChildsDiss}
A.~Childs.
\newblock {\em Quantum information processing in continuous time}.
\newblock PhD thesis, Massachusetts Institute of Technology, 2004.

\bibitem{Childs-2005-71}
{A.~M.} Childs, {D.~W.} Leung, and {M.~A.} Nielsen.
\newblock Unified derivations of measurement-based schemes for quantum
  computation.
\newblock {\em Phys. Rev. A}, 71:032318, 2005.

\bibitem{MessRechnen}
A.~Childs, D.~Leung, and M.~Nielsen.
\newblock Unified derivations of measurement-based schemes for quantum
  computation.
\newblock {\em arXiv:quant-ph/0404132}, 2004.

\bibitem{Cvetkovic}
D.~Cvetkovic.
\newblock {\em Eigenspaces of graphs}.
\newblock Cambridge University Press, 1997.

\bibitem{Da76}
E.~Davies.
\newblock {\em {Quantum theory of open systems}}.
\newblock Academic Press, London, 1976.

\bibitem{DiV}
D.~DiVincenzo.
\newblock Two-qubit gates are universal for quantum computation.
\newblock {\em Phys. Rev A}, 51:1015--1022, 1995.
 
\bibitem{Feynman:85}
R.~Feynman.
\newblock Quantum mechanical computers.
\newblock {\em Opt. News}, 11:11--46, 1985.

\bibitem{goldreich:promise}
O.~Goldreich.
\newblock On promise problems.
\newblock Technical Report~18, Electr. Colloquium Computational Complexity,
  2005.

\bibitem{Habil}
D.~Janzing.
\newblock {\em Computer Science Approach to Quantum Control}.
\newblock Uni-Verlag Karlsruhe, 2006.

\bibitem{ErgodicQutrits}
D.~Janzing.
\newblock {Spin-1/2 particles moving on a 2D lattice with nearest-neighbor
  interactions can realize an autonomous quantum computer}.
\newblock {\em Phys. Rev. A}, 75:012307, 2007.

\bibitem{Ergodic}
D.~Janzing and P.~Wocjan.
\newblock Ergodic quantum computing.
\newblock {\em Quant. Inf. Process.}, 4(2):129--158, 2005.

\bibitem{DiagonalEntryToC}
D.~Janzing and Wocjan P.
\newblock {A simple PromiseBQP matrix problem}.
\newblock {\em Theory of Computing}, 3:61--79, 2007.

\bibitem{RandomWalkBQP}
D.~Janzing and P.~Wocjan.
\newblock {BQP-complete problems concerning mixing properties of classical random walks on sparse graphs}.
\newblock arXiv:quant-ph/0610235.

\bibitem{BioBQP}
D.~Janzing and P.~Wocjan.
\newblock {A PromiseBQP-complete string rewriting problem}.
\newblock arXiv:quant-ph/0705.1180.

\bibitem{Ja68}
J.~M. Jauch.
\newblock {\em Foundations of quantum mechanics}.
\newblock Addison-Wesley, Reading, Mass., 1968.

\bibitem{KempeRegev}
J.~Kempe and O.~Regev.
\newblock 3-local Hamiltonian is QMA-complete.
\newblock {\em Quant. Inf. \& Comp.}, 3:258--264, 2003.

\bibitem{Kempe2local}
J.~Kempe, A.~Kitaev, and O.~Regev.
\newblock {The complexity of the local Hamiltonian problem}.
\newblock {\em Proc. 24th FSTTCS, accepted to SICOMP}, 2004.

\bibitem{KitaevShen}
A.~Kitaev, A.~Shen, and M.~Vyalyi.
\newblock {\em Classical and Quantum Computation}, volume~47.
\newblock Am. Math. Soc., Providence, Rhode Island, 2002.

\bibitem{Margolus:90}
N.~Margolus.
\newblock Parallel quantum computation.
\newblock In W.~Zurek, editor, {\em Complexity, Entropy, and the Physics of
  Information}. Addison Wesley Longman, 1990.

\bibitem{Shepherd}
D.~Shepherd, T.~Franz, and R.~Werner.
\newblock A universally programmable quantum cellular automaton.
\newblock {\em Phys. Rev. Lett.}, 97:020502, 2006.

\bibitem{RB00}
R.~Raussendorf and H.~Briegel.
\newblock Quantum computing via measurements only.
\newblock {\em Phys. Rev. Lett.}, page 5188, 2000.

\bibitem{Oliveira}
R.~Oliveira and B.~Terhal.
\newblock The complexity of quantum spin systems on a two-dimensional square
  lattice.
\newblock arXiv:quant-ph/0504050.

\bibitem{Trav}
B.~Travaglione and G.~Milburn.
\newblock
Generation of eigenstates using the phase-estimation algorithm.
\newblock {\em Phys. Rev. A}: 63, 032301, 2001.

\bibitem{Vollbrecht}
K.~Vollbrecht and I.~Cirac.
\newblock Quantum simulators, continuous-time automata, and translationally
  invariant systems.
\newblock arXiv:quant-ph/0704.3432.

\bibitem{WZ:06}
P.~Wocjan and S.~Zhang.
\newblock Several natural {BQP}-complete problems.
\newblock arXiv:quant-ph/0606179.

\bibitem{PSPACE}
P.~Wocjan, D.~Janzing, Th. Decker, and Th. Beth.
\newblock {Measuring 4-local $n$-qubit observables could probabilistically solve
  PSPACE}.
\newblock {\em Proceedings of the WISICT conference, Cancun 2004.} 
\newblock See also arXiv:quant-ph/0308011.


\end{thebibliography}
\end{document}